\def\mnras{{MNRAS}}

\newcommand{\etal}{\it et al.\rm}

\documentclass[cup5b]{caps}
\usepackage{graphicx}
\usepackage{amssymb}
\usepackage{ociwsymp3e}  

\HeadText{M. W. Bautz \& J. S. Arabadjis}  

\begin{document}

\pagenumbering{arabic}

\author[]{M. W. BAUTZ and J. S. ARABADJIS \\Massachusetts Institute of 
Technology Center for Space Research, Cambridge, MA, USA}

\chapter{The Dark Matter Distribution in \\Galaxy Cluster Cores}

\begin{abstract}

Galaxy cluster mass distributions  offer an important test of
the cold dark matter picture of structure formation,
and may even contain clues about the nature of dark matter.  
X-ray imaging spectroscopy of relaxed systems can
map cluster dark matter distributions, but are usually complicated by
the presence of central cool components  in the  intracluster medium.
Here we describe a statistically correct  approach  to distinguishing 
amongst simple alternative models of the cool component, and apply it to
one cluster. We also present mass profiles and central density slopes 
for five clusters derived from {\em Chandra} data, and illustrate 
how assumptions about the cool component affect the resulting mass profiles. 
For four of these objects, we find that the central density profile 
($r < 200 $h$_{50}^{-1}$ kpc) $\rho(r) \sim r^{\alpha}$ with $-2 < \alpha < -1$, for either of two models of the central cool component. These results are
consistent with standard CDM predictions.

\end{abstract}

\section{Introduction}

The cold dark matter (CDM) paradigm successfully describes many aspects of 
the formation of large-scale structure in the universe
(Lahav \etal\ 2001; Peacock \etal\ 2001; Navarro, Frenk \& White 1997;
Moore \etal\ 1999b).
Galaxy-scale dark matter halos, 
however, exhibit several
apparent inconsistencies with CDM, for example:  the number of Milky Way
satellites appears to be at least an order of magnitude lower than CDM
predictions 
(Kauffman, White, \& Guideroni 1993; Moore \etal\ 1999a; Klypin \etal\ 1999),
and dark matter halos in dwarf and low surface
brightness galaxies are much less cuspy than in CDM simulations 
(Burkert 1995; McGaugh \& de Blok 1998; Moore \etal\ 1999b).
Some reports
(Tyson, Kochanski \& Dell'Antonio 1998; Smail \etal\ 2000)
even suggest that CDM fails on galaxy cluster
scales for some clusters, but the latter are controversial 
(Broadhurst \etal\ 2000; Shapiro \& Iliev 2000).

Proposed modifications of CDM include self-interacting dark matter 
(Spergel \& Steinhardt 2000; Firmani \etal\ 2000),
warm dark matter 
(Hogan \& Dalcanton 2000),
annihilating dark matter 
(Kaplinghat, Knox \& Turner 2000),
scalar field dark matter 
(Hu \& Peebles 2000; Goodman 2000),
and mirror matter
(Mohapatra, Nussinov \& Teplitz 2002),
each of which is invoked to soften the core density profile.
Many of these modifications will soften the core profile of galaxy clusters as
well, although other astrophysical processes such as the adiabatic contraction
of core baryons 
(Hennawi \& Ostriker 2002)
may ameliorate this effect.  Baryons, however,
introduce a host of complications to CDM simulations 
(Frenk 2002).

In order better to discriminate among CDM, its modifications, and other 
astrophysical influences, we are mapping the dark matter profiles of a
sample of galaxy cluster cores.  We use imaging spectroscopy from the
{\it Chandra X-ray Observatory}  (our own observations --
 Arabadjis, Bautz \& Garmire 2002 --
and those in the {\em Chandra} archive)
to deproject the  radial profile of the intracluster medium (ICM) density
and temperature for each cluster.  
In order to extract a dark matter profile
from spatially resolved X-ray spectroscopy one usually assumes that the galaxy
cluster is spherically symmetric and in hydrostatic equilibrium, and so for the
most part we have restricted our sample to clusters for which, to judge from
their {\em Chandra} X-ray images,  these assumptions
appear to be valid.  As noted by Allen (1998), clusters in hydrostatic
equilibrium often 
contain ``cooling flows,'' or cool components in their cores, and these
components must be properly modelled if reliable inferences about cluster
mass are to be drawn.
If the model of the ICM contains only a single
emitting component (at temperature $T$ and density $\rho$) at each radius, the
inferred temperature profile will tend to dip significantly toward the center
of a cluster with a cool component.  
If, however, gas in the the core 
contains a second (cooler) component which is cospatial and isobaric with the
first, then a hot ICM component coexists with the cool component in the 
core.  The latter
case tends to produce a larger central mass (see Figures~\ref{fig1})
and steeper density profile than the  former.  

In this contribution, we first discuss the proper method for distinguishing
between these two simple models, and then present results for mass profiles
for a number of clusters.

\section{One Temperature or Two?}
\subsection{Approach}
Our problem is to 
choose between a simple model {\sf M}$^{\sf s}$, representing a 
single-com-
ponent core ICM,  and a complex model {\sf M}$^{\sf c}$, 
representing two-phase gas.
In order to test for the presence of a second emission component we adopt a
simplified geometry containing only two spherical shells (inner = 1,
outer = 2).  In both models ({\sf M}$^{\sf s}$ and {\sf M}$^{\sf c}$), shell 2
contains a (hot) thermal plasma at temperature $T_{2h}$ and density
$\rho_{2h}$.  Model {\sf M}$^{\sf s}$ contains only one emission component in
shell 1, characterized by a temperature $T_{1h}$ and a density $\rho_{1h}$,
whereas model {\sf M}$^{\sf c}$ contains a hot and a cool emission component in
shell 1, described by $T_{1h}$, $\rho_{1h}$, $T_{1c}$, and $\rho_{1c}$.
The X-ray emission from each component is modelled spectroscopically using
the {\sc Mekal} model (see, e.g., 
Liehdal, Osterheld\ \& Goldstein 1995)
as implemented in the {\sc Xspec} software package 
(Arnaud 1996).
The best-fit parameter values of each model are
calculated using a $\chi^2$ minimization routine.  Hereafter we will refer to
the simple and complex model parameters using vectors $\theta^{\sf s}$ and
$\theta^{\sf c}$, respectively; i.e.,
$\theta^{\sf s}=(T_{1h},\rho_{1h},T_{2h},\rho_{2h})$ and
$\theta^{\sf c}=(T_{1h},\rho_{1h},T_{1c},\rho_{1c},T_{2h},\rho_{2h})$.

It might seem straightforward to apply a conventional test such as
the likelihood ratio test or the
$F$-test 
(Bevington 1969; Cash 1979)
to choose between the two models.  However, since $\theta^{\sf s}$
lies on a boundary of $\theta^{\sf c}$ (with $\rho_{1c}=0$), these tests
cannot be employed 
(Protassov \etal\ 2002).
Instead, we construct an {\it empirical}
$F$-distribution using Markov Chain Monte Carlo ({\sc mcmc}) sampling, and
gauge the significance of the complex model from the location of the $F$ value
of the data within that distribution 
(Protassov \etal\ 2002).

\subsection{An empirical ``$F$-distribution''}

We briefly summarize our computation of an empirical ``$F$-distribution'' here;
details will be presented elsewhere (Arabadjis \& Bautz, in preparation).
Starting with the best-fit parameters $\theta_0$ of model {\sf M}$^{\sf s}$, we
sample the 4D parameter space in its vicinity using {\sc mcmc} sampling.  This
is done by running a Tcl script within {\sc Xspec} which calculates the
probability distribution function $P$ of a trial perturbation $\theta_1$ about
$\theta_0$ given the observed data.  The trial point is chosen using the trial
distribution function $q(\theta_0,\theta_1)$.  The choice of $q$ is arbitrary;
we restrict ourselves to functions which are symmetric in parameter space
transitions, i.e.\ $q(\theta_i,\theta_j) = q(\theta_j,\theta_i)$. 
This new parameter set is accepted if
$P(\theta_1)/P(\theta_0)$ exceeds a random number on [0,1].  If not, the trial
point is rejected and new one is selected.  The sequence of accepted $\theta_i$
is a Markov Chain whose stationary distribution follows $P(\theta)$ 
(Lewis \& Bridle 2002).
We repeat this procedure until we have 100 values of $\theta$ for model
{\sf M}$^{\sf s}$.

\begin{figure}
  \includegraphics[height=160pt]{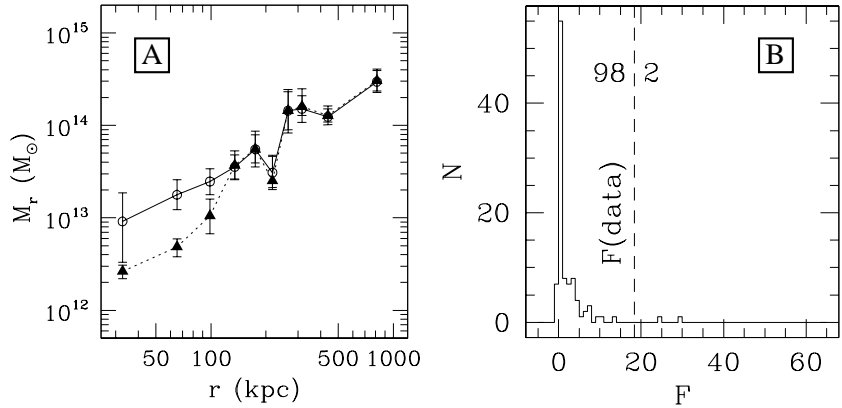}
  \caption{[A] One- and two-temperature mass profiles (dotted/triangles and
solid/circles, respectively) of EMSS 1358+6245; [B] empirical $F$-distribution
for models ${\sf M}^{\sf s}$ and ${\sf M}^{\sf c}$ of EMSS 1358+6245. \label{fig1}}
\end{figure}

For each of the parameter sets in the sample we simulate an X-ray spectrum,
taking proper account of the instrument response and photon statistics.  We
then model each of the simulated data sets using both {\sf M}$^{\sf s}$ and
{\sf M}$^{\sf c}$, and tabulate the $F$-value of each data set:

\begin{equation}
F = \frac{\chi^2({\theta}^{\sf s})-\chi^2({\theta}^{\sf c})}
{\chi^2({\theta}^{\sf s})/\nu({\sf M}^{\sf s})}
\end{equation}

\noindent where $\nu({\sf M}^{\sf s})$ is the number of degrees of freedom of
the simple model.  (In practice, the normalization can be ignored.)  The
$F$-distribution for the cooling flow cluster EMSS 1358+6245 is shown in
panel B of Figure~\ref{fig1}.  The $F$-value of the original {\it Chandra} data set is indicated
with a dashed line.

\subsection{An example: MS1358+6245}
Of the 100 {\sc mcmc} simulations that were run, only two resulted in an
$F$-value which exceeded that of the data -- that is, if ${\sf M}^{\sf s}$ were
the correct description, an $F$-value as large as that observed would occur
with a probability of only 2\% -- meaning that the model with a separate,
co-spatial cool component is preferred.  The result is that a model with a
steeper density profile and a larger central mass is favored.  If this trend
obtains for most cooling flow clusters, it may rule out several of the CDM
modifications.

\section{Mass profiles and core slopes}
\begin{figure}
{\centering \parbox{0.46\columnwidth}{
\includegraphics[width=0.45\columnwidth]{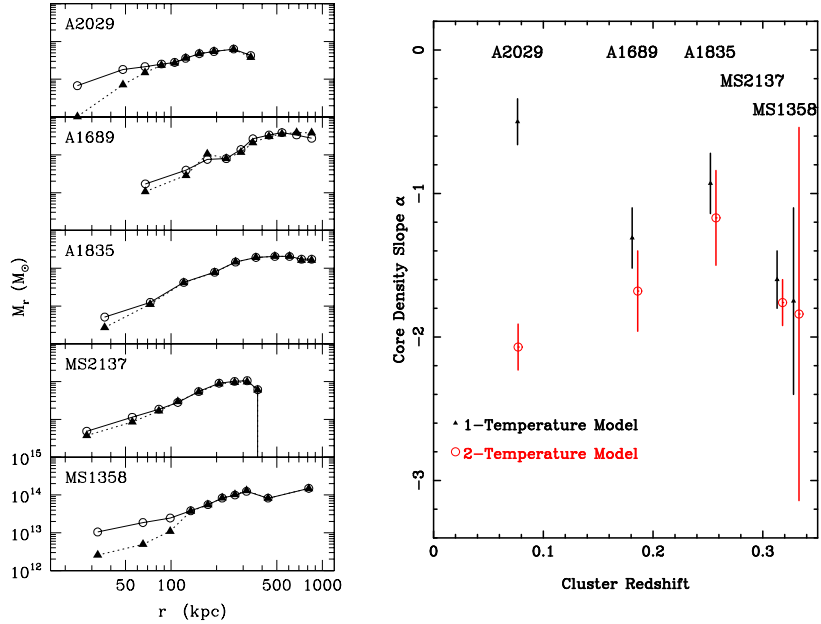}
} \hfil 
\parbox{0.50\columnwidth}{
\includegraphics[width=.49\columnwidth]{index3_pretty.ps}
}
}
\caption{ {\em Left: } {\em Chandra} mass profiles,  
assuming 1-temperature and  2-temperature central emission models, 
(dotted/triangles and solid/circles, respectively) for 5 clusters.
Masses and radii for $H_{0} = 0, \Omega_{m} = 1$. 
{\em Right: } Core density
power-law index $\alpha$ (for $\rho(r) \sim r^{\alpha}$ ) determined 
within $r < 200$ h$_{50}^{-1}$ kpc for the same clusters, using 
1- and 2-temperature models (open circles \& filled triangles, respectively.)
The error bars show 90\%-confidence intervals for $\alpha$.}
\label{fig2}
\end{figure}

We have yet to apply the foregoing analysis to discriminate between 
1- and 2-temperature models to all clusters in our sample. Instead,
here we merely illustrate, in  Figure~\ref{fig2}, 
the sensitivity of the derived mass profile and inferred density slope in 
the cluster core to the assumed state of the intracluster medium for five
reasonably relaxed clusters.  These
results have been obtained using the deprojection methods described
in Arabadjis, Bautz \& Garmire (2002). 
The left panel of Figure~\ref{fig2} shows that, generally,
the inferred mass is larger, the encircled mass profile 
marginally flatter, and, in consequence, the density somewhat steeper 
in the 2-temperature model.

The right panel of Figure~\ref{fig2} shows the logarithmic density slope
parameter $\alpha$ (here defined so that the central density 
$\rho(r) \sim r^{\alpha}$) inferred from the mass profile for each
cluster. (We plot $\alpha$ against redshift merely as a convenient means of
display.)  The indices shown
were obtained by fitting the mass profile  within $r < 200$ h$_{50}^{-1}$ kpc, 
which is typically about one-tenth the virial radius inferred from the best-fit
NFW profiles for these objects.  
We note that the core slopes are generally consistent
with, or slight steeper than, the standard CDM results ($-1.5 < \alpha < -1.0$;
Navarro, Frenk \& White 1997; Moore \etal\ 1999b).

\section{Summary}
We have illustrated a  rigorous and quantitative
procedure for distinguishing between emission models of cool components 
in clusters. 
We have presented mass profiles derived from {\em Chandra} X-ray images for
five clusters, using both 1- and 2-temperature models for the central X-ray
emission. In almost all cases we find that the slope of the total 
central mass density is consistent with, or slightly steeper than, 
standard CDM predictions.

\newcommand{\ARAA}[2]{\it ARA\&A,\bf\ #1,\rm\ #2}
\newcommand{\ApJ}[2]{\it ApJ,\bf\ #1,\rm\ #2}
\newcommand{\ApJL}[2]{\it ApJL,\bf\ #1,\rm\ #2}
\newcommand{\ApJSS}[2]{\it ApJS,\bf\ #1,\rm\ #2}
\newcommand{\AandA}[2]{\it A\&A,\bf\ #1,\rm\ #2}
\newcommand{\AandASS}[2]{\it A\&AS,\bf\ #1,\rm\ #2}
\newcommand{\AJ}[2]{\it AJ,\bf\ #1,\rm\ #2}
\newcommand{\BAAS}[2]{\it BAAS,\bf\ #1,\rm\ #2}
\newcommand{\ASP}[2]{\it ASP Conf.\ Ser.,\bf\ #1,\rm\ #2}
\newcommand{\JCP}[2]{\it J.\ Comp.\ Phys.,\bf\ #1,\rm\ #2}
\newcommand{\MNRAS}[2]{\it MNRAS,\bf\ #1,\rm\ #2}
\newcommand{\N}[2]{\it Nature,\bf\ #1,\rm\ #2}
\newcommand{\PASJ}[2]{\it PASJ,\bf\ #1,\rm\ #2}
\newcommand{\PASP}[2]{\it PASP,\bf\ #1,\rm\ #2}
\newcommand{\PRD}[2]{\it Phys.\ Rev.\ D,\bf\ #1,\rm\ #2}
\newcommand{\PRL}[2]{\it Phys.\ Rev.\ Lett.,\bf\ #1,\rm\ #2}
\newcommand{\RPP}[2]{\it Rep.\ Prog.\ Phys.,\bf\ #1,\rm\ #2}
\newcommand{\ZA}[2]{\it Z.\ Astrophs.,\bf\ #1,\rm\ #2}
%

\end{document}